\documentclass[%
 aip,
 amsmath,amssymb,
 reprint,%
]{revtex4-1}

\usepackage{graphicx}
\usepackage{dcolumn}
\usepackage{bm}

\usepackage[utf8]{inputenc}
\usepackage[T1]{fontenc}
\usepackage{mathptmx}
\usepackage{etoolbox}

\makeatletter
\def\@email#1#2{%
 \endgroup
 \patchcmd{\titleblock@produce}
  {\frontmatter@RRAPformat}
  {\frontmatter@RRAPformat{\produce@RRAP{*#1\href{mailto:#2}{#2}}}\frontmatter@RRAPformat}
  {}{}
}%
\makeatother
\begin{document}

\preprint{AIP/123-QED}

\title{Cubic magneto-optic Kerr effect in Co(111) thin films}
\author{M. Gaerner}
\affiliation{Faculty of Physics, Bielefeld University, 33615 Bielefeld, Germany}
 \email{mgaerner@physik.uni-bielefeld.de.}
\author{R. Silber}%
\affiliation{Department of Materials Engineering and Recycling, Faculty of Materials Science and Technology, VSB-Technical University of Ostrava, Ostrava 70800, Czech Republic}
\affiliation{Nanotechnology Centre, Centre for Energy and Environmental Technologies, VSB-Technical University of Ostrava, Ostrava 70800, Czech Republic}

\author{M. Schäffer}
\affiliation{Faculty of Physics, Bielefeld University, 33615 Bielefeld, Germany}
\author{J. Hamrle}
\affiliation{Faculty of Mathematics and Physics, Charles University, Prague 12116, Czech Republic}
\affiliation{Faculty of Nuclear Sciences and Physical Engineering,
Czech Technical University, Prague 12000, Czech Republic}
\author{A. Ehrmann}
\affiliation{%
Faculty of Engineering and Mathematics, Bielefeld University of Applied Sciences and Arts, 33619 Bielefeld, Germany}
\author{M. Wortmann}
\affiliation{Faculty of Physics, Bielefeld University, 33615 Bielefeld, Germany}
\affiliation{%
Faculty of Engineering and Mathematics, Bielefeld University of Applied Sciences and Arts, 33619 Bielefeld, Germany}
\author{T. Kuschel}
\affiliation{Faculty of Physics, Bielefeld University, 33615 Bielefeld, Germany}
\affiliation{Institute of Physics, Johannes Gutenberg University Mainz, 55128 Mainz, Germany}
\date{\today}

\begin{abstract}
The magneto-optic Kerr effect (MOKE) is often applied as a tool for the magnetic characterization of thin films. Here, the change in polarization upon reflection from the magnetized sample is mainly regarded as being linearly proportional to the magnetization $\bm{M}$ (LinMOKE). MOKE contributions of second order in $\bm{M}$, also known as quadratic MOKE (QMOKE), which are proportional to $\bm{M}^2$, have also been studied in the past and used in thin film characterization. Recently, we reported on a systematic investigation of third-order MOKE contributions, named cubic MOKE (CMOKE) in Ni(111) thin films. This CMOKE manifests itself as an anisotropic contribution to the MOKE signal (with regard to the crystallographic orientation) measured in longitudinal or transversal configuration in full magnetic saturation. While LinMOKE (odd in $\bm{M}$) and QMOKE (even in $\bm{M}$) can easily be separated by methods based on magnetization parity, this no longer holds true for LinMOKE and CMOKE (odd in $\bm{M}$). It is therefore crucial to be aware of CMOKE contributions in order to correctly interpret MOKE data. Here, we report on the observation of CMOKE in thin film heterostructures with structurally twinned and untwinned Co(111) layers, demonstrating that a large CMOKE is not only present in Ni thin films. Additionally, we show that the observed anisotropic contributions cannot stem from LinMOKE by analyzing their dependence on the angle of incidence (AoI) of light. While the QMOKE is almost vanishing in Co(111) using light with wavelengths of 635\,nm and 406\,nm, the CMOKE contributions reach up to about 30\% of the LinMOKE contribution at an AoI of 45 degrees and become even more dominant towards normal AoI, which emphasizes the importance of higher-order MOKE effects in magneto-optic experiments.
\end{abstract}

\maketitle

\begin{figure*}[]
\begin{center}
\includegraphics[scale=0.21]{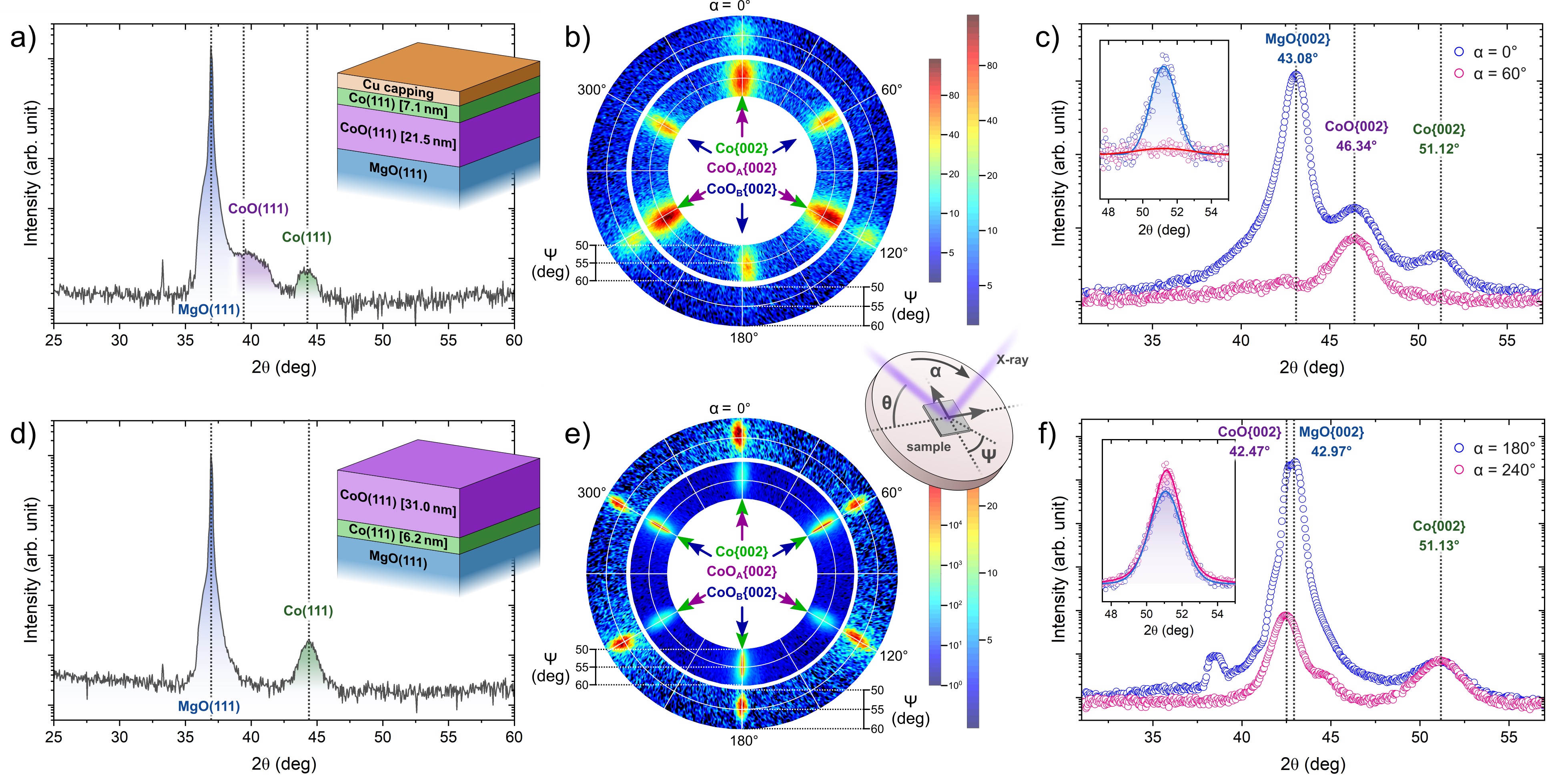}
\end{center}
\vspace{-7mm}
\caption{(a,d) Specular XRD with insets showing the respective layer stacks, (b,e) off-specular texture maps of the CoO\{002\} (inner ring) and Co\{002\} (outer ring) diffraction peaks and (c,f) corresponding off-specular $\theta $-2$\theta $ scans for samples 1 and 2, respectively. Insets in (c,f) show the Co\{002\} peaks at $\alpha $=0$^\circ $,60$^\circ $ after baseline subtraction. The inset in (e) shows the experimental geometry for off-specular measurements.}
\vspace{-3mm}
\label{XRD}
\end{figure*}
\noindent The magneto-optic Kerr effect (MOKE) \cite{JohnKerrLL.D..1877} serves as a powerful tool in the area of thin-film magnetic characterization \cite{E.R.Moog.1985, Qiu.2000}. It describes the change of light polarization upon reflection from a magnetized sample based on magnetic circular dichroism and magnetic circular birefringence during reflection. Most studies employing MOKE rely on the assumption that the polarization change of light depends linearly on the magnetization $\bm{M}$ \cite{Florczak.1990, Vavassori.2000, Harp.1993, Veis.2014, Schafer.2007, JeffreyMcCord.2015, Beaurepaire.1996, Kirilyuk.2010}. In most cases, linear MOKE (LinMOKE) can be separated from quadratic MOKE (QMOKE)\cite{Metzger.1965, JHamrle.2007}, being $\propto \bm{M}^2$, by magnetization curve symmetrization since LinMOKE and QMOKE are odd and even with respect to $\bm{M}$, respectively. QMOKE has been utilized to sense the structural ordering in Heusler compounds \cite{Wolf.2011, Silber.2020}, to develop advanced vectorial magnetrometry techniques \cite{Mewes.2004, TKuschel.2011, Kuschel.2012, Tesarova.2012, Fan.2016, Jang.2020} as well as QMOKE microscopy \cite{Janda.2018, Xu.2019} and spectroscopy \cite{Silber.2018, Silber.2019, Wohlrath.2025}, to investigate spin-orbit torques magneto-optically \cite{Tesarova.2013, Fan.2014, Montazeri.2015} and to study antiferromagnetic materials \cite{Saidl.2017, Zheng.2018, Zhao.2021}. Second order MOKE originating from non-uniform $\bm{M}$ became known as Schäfer-Hubert effect \cite{Schafer.1990, Kambersky.1992, Schafer.1995}.\\
MOKE of third order in $\bm{M}$ has rarely been mentioned in literature in the past \cite{Petukhov.1998, Gridnev.1997, K.Postava.2004}. Only recently, the cubic magneto-optic Kerr effect (CMOKE) was investigated in Ni(111) thin films \cite{Gaerner.2024} as a first systematic study. The MOKE signal that is measured in a longitudinal MOKE (LMOKE) geometry (magnetic field applied in-plane with the sample surface and the plane of incidence) was shown to be proportional not only to the magnetization component along the magnetic field $\bm{M}_L$, but also to $\bm{M}_L^3$. While the LMOKE contribution $\propto \bm{M}_L$ is isotropic in its dependence on the sample orientation in the case of cubic crystal structures, this is not the case for the longitudinal CMOKE (LCMOKE) contribution $\propto \bm{M}_L^3$. For (111)-oriented cubic crystals, the CMOKE manifests itself as a threefold angular dependence of the saturated MOKE signal on the in-plane sample rotation \cite{Gaerner.2024}. Just as LinMOKE, CMOKE is odd with respect to $\bm{M}$ and can thus not be separated from LinMOKE by magnetization curve symmetrization. Therefore, it is important to be aware of possible CMOKE contributions when analyzing MOKE data. One advantage of CMOKE over LinMOKE for (111)-oriented cubic crystals is the sensitivity to in-plane magnetization for perpendicular incidence of light \cite{Gaerner.2024, Pan.2024}.\\
Here, we show that a large CMOKE can also be detected in Co(111) thin films, proving that the effect is not unique to Ni films. Furthermore, we analyze the dependence of the CMOKE signal on the structural domain twinning. Therefore, heterostructures containing Co(111) and CoO(111) layers have been grown on MgO(111) substrates using molecular beam epitaxy (MBE). The degree of structural domain twinning in the Co(111) layer then depends on its position in the stacking order of the heterostructure. Co(111) grown directly on MgO(111) is structurally twinned while Co(111) grown on CoO(111) shows no twinning. Such heterostructures are archetype systems for the investigation of exchange bias \cite{Meiklejohn.1956, Meiklejohn.1957} and are investigated up to this day \cite{Wortmann.2023, OliveiraSilva.2025}, also commonly using MOKE. This underlines the importance of CMOKE investigations in such a system, especially when the CMOKE reaches 30\% of the LinMOKE as observed here. Additionally, we investigate the dependence of the CMOKE contribution on the angle of incidence (AoI) of light and compare it to the theoretical prediction made by Yeh's 4x4 matrix formalism. Thus, we show that the anisotropic contribution cannot stem from LinMOKE. \\ \\
\noindent The permittivity tensor of a magnetized crystal $\bm{\varepsilon }$ can be described up to third order in $\bm{M}$ using the Taylor expansion
\begin{align}
    \bm{\varepsilon } (\bm{M}) = \bm{\varepsilon } ^{(0)} + \bm{\varepsilon } ^{(1)} + \bm{\varepsilon } ^{(2)} + \bm{\varepsilon } ^{(3)}
\end{align}
in which the superscript denotes the order of dependence on $\bm{M}$. Using the Einstein summation, the elements of $\bm{\varepsilon }$ can then be expressed as
\begin{align}
	\varepsilon_{ij}&=\varepsilon_{ij}^{(0)}  +  K_{ijk}M_k  +  G_{ijkl} M_k M_l + H_{ijklm} M_k M_l M_m \text{ .} 
\label{Perm_3rd_order}
\end{align}
Here, $K_{ijk}$, $G_{ijkl}$ and $H_{ijklm}$ are the elements of the linear, quadratic and cubic magneto-optic (MO) tensors $\bm{K}$, $\bm{G}$ \cite{Visnovsky1986, JHamrle.2007, TKuschel.2011} and $\bm{H}$ \cite{K.Postava.2004, Gaerner.2024} while $M_{k/l/m}$ are the components of the magnetization vector $\bm{M}=[M_x, M_y, M_z]=[M_T, M_L, M_P]$ with transverse ($M_T$), longitudinal ($M_L$) and polar ($M_P$) magnetization. Inserting this expression into the analytical approximation for MOKE in thin ferromagnetic layers \cite{Hamrle.2003, Visnovsky.2018} results in an equation for the Kerr angles which entails contributions with different dependencies on $\bm{M}$. Most of these contributions can be experimentally separated by using the eight-directional method \cite{Postava.2002}, in which the MOKE signal is measured for eight different in-plane magnetization directions $\mu = k\cdot 45^\circ , k=\{0,1, ..., 7\}$. The signals for different directions are then added or subtracted from each other to separate MOKE contributions with different dependencies on $\bm{M}$. For details on separation techniques using the eight-directional method, see Refs. ~\onlinecite{Postava.2002, Silber.2019, Gaerner.2024}. For (111)-oriented cubic crystals, the MOKE contribution for the magnetically saturated $M_L$ component can then be written as \cite{Gaerner.2024, Silber2026}
\begin{align}
    \Phi _{M_L, M_L^3}^{(111)} &= \frac{1}{2}\left( \Phi _{s/p}^{\mu = 90^\circ } - \Phi _{s/p}^{\mu = 270^\circ } \right) \nonumber \\
    & = \pm B_{s/p} \left(K + \frac{H_{123}+3H_{125}}{2} \right) \nonumber \\
    & \quad -A_{s/p} \frac{\sqrt{2}}{6}\left( \Delta H + \frac{K\Delta G}{\varepsilon _d } \right) \sin{(3\alpha)} \text{  .}
    \label{ML}
\end{align}
Here, $\varepsilon _d$ represents the non-magnetic diagonal permittivity tensor elements. $K$ is the MO parameter of first order in $\bm{M}$ while $\Delta G=G_{11}-G_{12}-2G_{44}$ and $\Delta H=H_{123}-3H_{125}$ are anisotropy parameters which describe the anisotropic strengths of $\bm{G}$ and $\bm{H}$, respectively. We can identify an isotropic term that depends on the optical weighting factor $B_{s/p}$ that is maximal near grazing incidence and zero for perpendicular incidence. Additionally, an anisotropic threefold term depending on the sample angle $\alpha $, purely $\propto M_L^3$, can be identified, which depends on the optical weighting factor $A_{s/p}$. While $B_{s/p}$ is odd with regard to the AoI, $A_{s/p}$ is even, so that it has a finite value at normal AoI.
\begin{figure*}
\begin{center}
\includegraphics[scale=0.175]{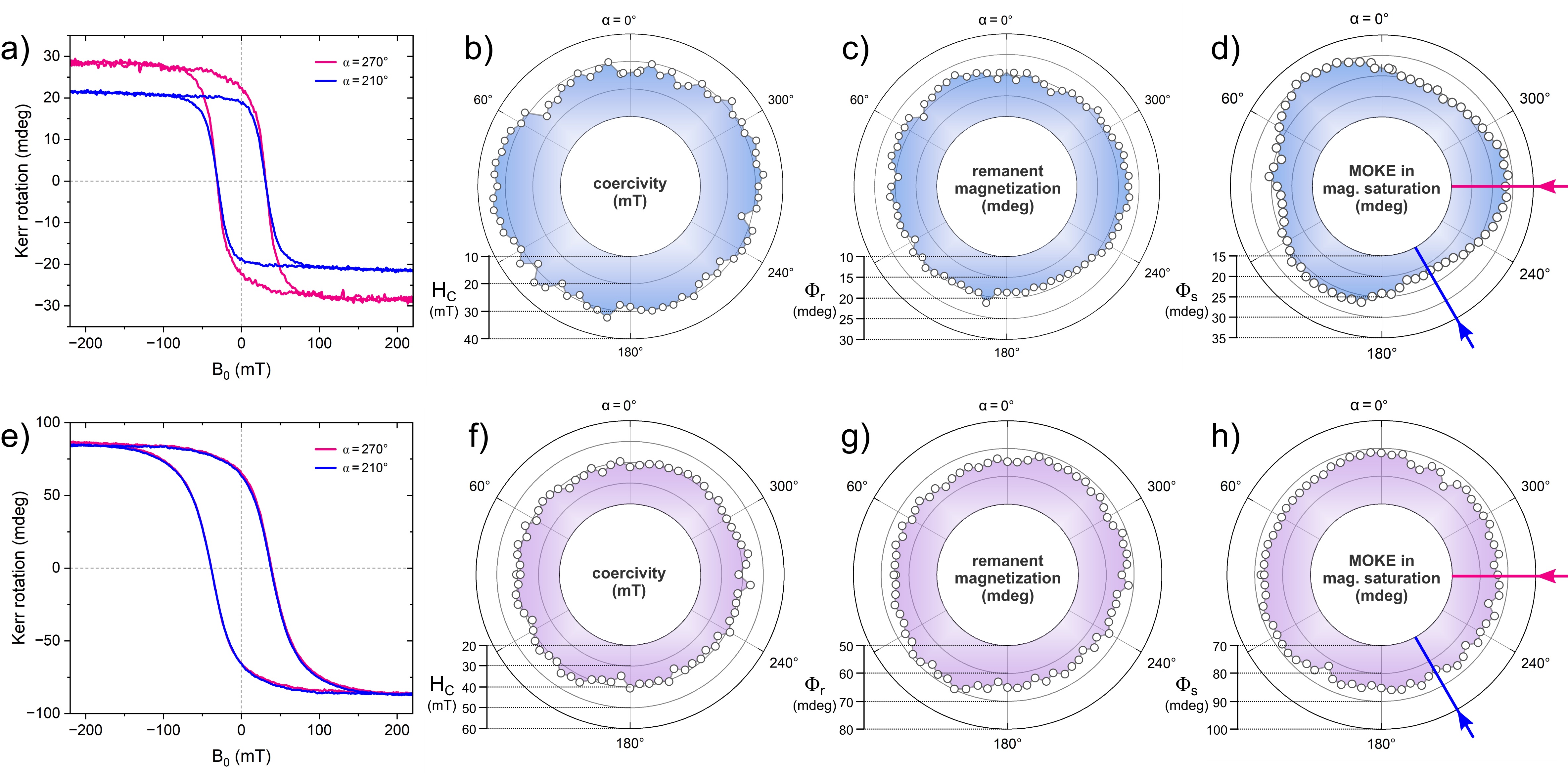}
\end{center}
\vspace{-7mm}
\caption{(a,e) MOKE curves at sample positions of $\alpha $=210$^\circ $ and $\alpha $=270$^\circ $, (b,f) coercive field and (c,g) remanent magnetization depending on $\alpha $, (d,h) MOKE in magnetic saturation at $B_0$=230\,mT depending on $\alpha $ for samples 1 and 2, respectively.}
\vspace{-3mm}
\label{MOKE}
\end{figure*}
\newpage
\noindent The transversal MOKE (TCMOKE) signal of the magnetization component $M_T$
\begin{align}
    \Phi _{M_T^3}^{(111)} &= \frac{1}{2}\left( \Phi _{s/p}^{\mu = 0^\circ } - \Phi _{s/p}^{\mu = 180^\circ } \right) \nonumber \\
    & = A_{s/p} \frac{\sqrt{2}}{6}\left( \Delta H + \frac{K\Delta G }{\varepsilon _d } \right) \cos{(3\alpha)}
    \label{MT}
\end{align}
is purely $\propto M_T^3$ and possesses an anisotropic threefold dependence on $\alpha $ \cite{Gaerner.2024, Silber2026}. Furthermore, the eight-directional method is also able to separate the QMOKE components $\Phi _{M_LM_T} \propto \sin{(3\alpha )}$ and $\Phi _{M_T^2-M_L^2} \propto \cos{(3\alpha )}$ (see supplemental material). \\ \\
\noindent The samples presented in this letter have been grown on MgO(111) substrates via MBE (see supplemental material for details). For the primary sample 1, a CoO buffer layer of 21.5\,nm thickness has been deposited. Afterwards, a 7.1\,nm thick Co layer was grown on the CoO and capped with Cu. The structural characterization using X-ray diffraction (XRD) and X-ray reflectivity (XRR) was carried out using the Cu $K_\alpha $ source of a Phillips X'pert Pro MPD PW3040-60 diffractometer. Figure \ref{XRD}(a) shows the specular XRD pattern of sample 1. Both the Co(111) and CoO(111) peaks are clearly visible and indicate crystalline growth of fcc-type Co and the NaCl-type CoO.
The 2$\theta $ value of the Co(002) peak corresponds to an out-of-plane (oop) lattice constant of 3.56\,\AA, close to the bulk value of 3.54\,\AA \cite{Owen.1954, LaPenaOShea.2010}. The CoO(002) 2$\theta $ peak position corresponds to an oop lattice constant of 3.94\,\AA, which differs from the bulk value of 4.25\,\AA \cite{Tran.2006}, indicating a strain of 7.9\% in the oop direction. \\
In Fig. \ref{XRD}(b), off-specular XRD Euler's cradle texture maps of the Co\{002\} and CoO\{002\} peaks are shown. All texture maps were measured for 360$^\circ $ of sample rotation $\alpha $ with a sample tilt of $\Psi =\langle 50 ^\circ $, $60^\circ \rangle $. The CoO\{002\} and Co\{002\} peaks were then captured at 2$\theta $=46.582$^\circ $ and 2$\theta $=51.304$^\circ $, respectively. Each structural domain orientation of the thin films can be identified by a threefold diffraction peak pattern. The sixfold symmetry, i.e. two threefold diffraction peak patterns with $\Delta \alpha $=60$^\circ $ in-plane rotation, of the observed CoO\{002\} peaks shows that twinning is present in the CoO-layer. This twinning manifests itself in two sublattices with a $\Delta \alpha $=60$^\circ $ in-plane rotation relative to each other due to incompatible stacking orders (ABC and ACB). In contrast, the texture map for the Co(111) layer shows only one dominant threefold symmetry, indicating that the Co layer is only comprised of one dominant structural phase. To quantitatively determine the degree of twinning in the Co(111) layer (analogously to Ref. ~\onlinecite{Gaerner.2024}), $\theta$-2$\theta $ scans of the Co\{002\} diffraction peaks at $\alpha $=0$^\circ $ and $\alpha $=60$^\circ $, corresponding to the different structural twinning phases, have been made and are shown in Fig. \ref{XRD}(c). The peaks have been fitted using Pseudo-Voigt-functions after baseline subtraction utilizing an exponential decay function. As no visible Co\{002\} peak can be identified at $\alpha $=60$^\circ$, the Co-layer can be regarded fully untwinned (c.f. inset of Fig. \ref{XRD}(c)). \\
\noindent For sample 2, a Co layer of 6.2\,nm thickness was directly deposited onto the MgO(111) substrate. Subsequently, a CoO layer of 31.0\,nm thickness was grown on top. The specular XRD pattern of sample 2 is shown in Fig. \ref{XRD}(d). Again, the Co(111) peak indicates good crystalline growth of the Co  with an oop lattice constant of 3.54\,\AA. The CoO(111) peak is not visible, as it overlaps with the MgO(111) peak when the CoO is not strained ($a_{MgO}=4.21$\AA). The off-specular texture maps for sample 2, presented in Fig. \ref{XRD}(e), show that both the Co\{002\} and CoO\{002\} peaks possess sixfold symmetries and can therefore both be considered twinned. Using the $\theta$-2$\theta $ scans of the Co\{002\} peaks at $\alpha $=180$^\circ $ and $\alpha $=240$^\circ $ after baseline subtraction (see the inset of Fig. \ref{XRD}(f)), the degree of twinning in the Co(111) layer can be determined by comparing the areas under the fitted peaks for the two different structural phases \cite{Gaerner.2024}. This results in a degree of twinning of 86.9\%.
\begin{figure}
\begin{center}
\includegraphics[scale=0.125]{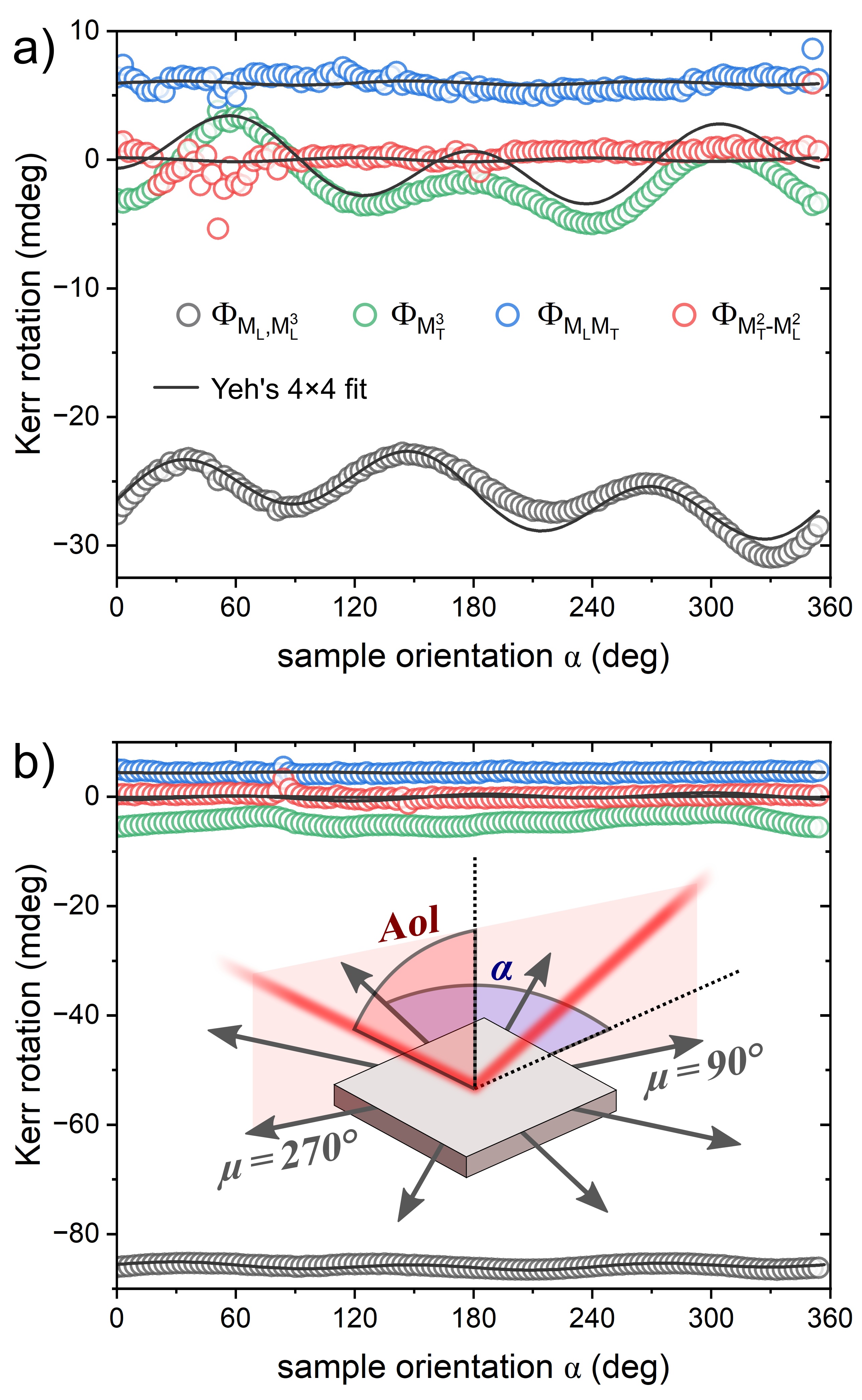}
\end{center}
\vspace{-7mm}
\caption{Results of the eight-directional measurements for (a) sample 1 and (b) sample 2 including fits of the numerical simulations. The finite offset of $\Phi _{M_T^3}$ cannot be simulated. Therefore, the fits of $\Phi _{M_T^3}$ and $\Phi _{M_T^2-M_L^2}$ overlap for sample 2. The inset shows the experimental geometry. Magnetic field directions $\mu = 90^\circ /270^\circ $ are parallel to the plane of incidence.}
\vspace{-3mm}
\label{8dir}
\end{figure}

\begin{figure}[h!]
\begin{center}
\includegraphics[scale=0.245]{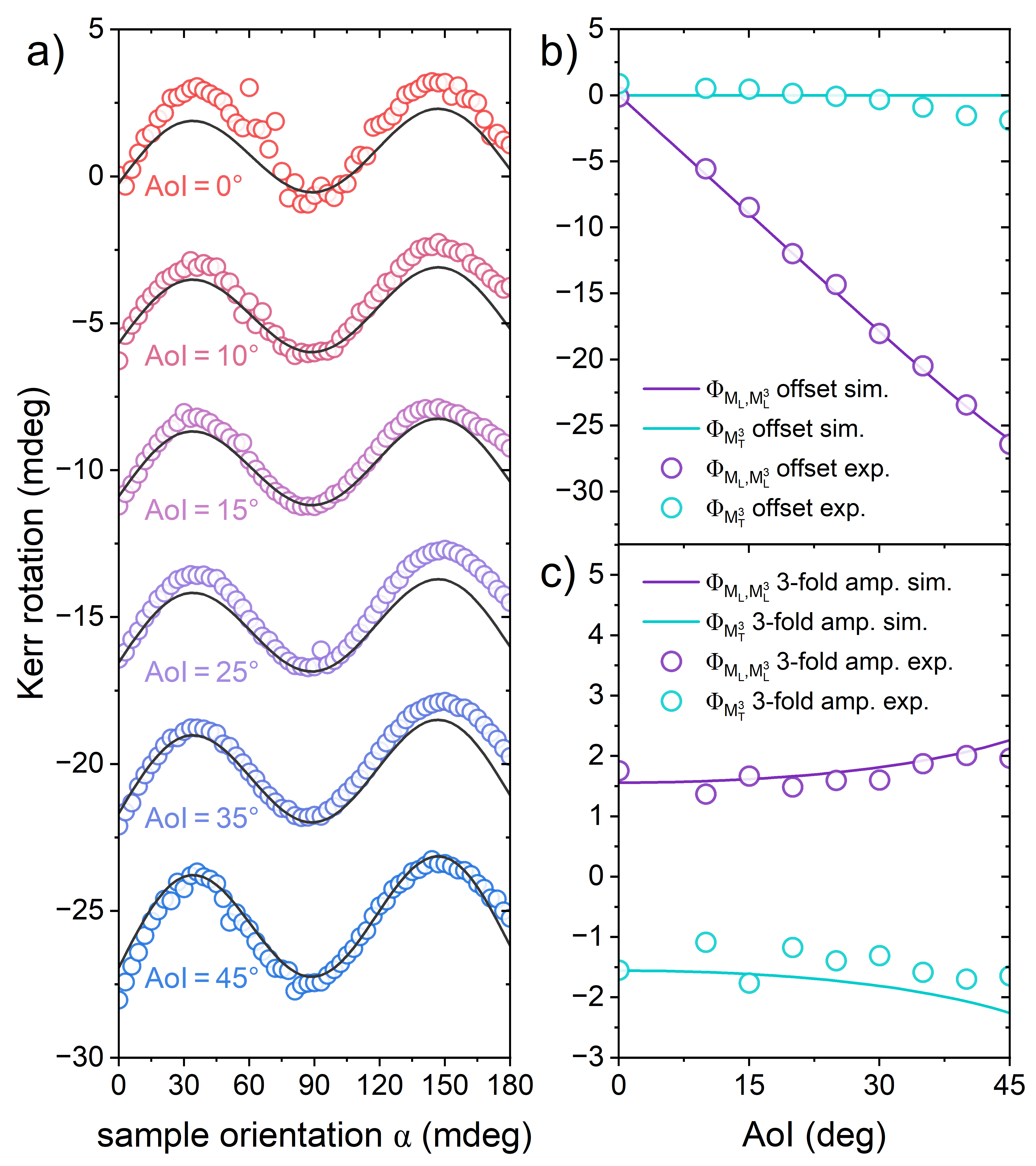}
\end{center}
\vspace{-7mm}
\caption{Measurement of sample 1 (circles) and simulations (solid black lines) of (a) the angular dependence of $\Phi _{M_L,M_L^3}$ with varying AoI, (b) offsets and (c) amplitudes of $\Phi _{M_L,M_L^3}$ and $\Phi _{M_T^3}$  depending on the AoI.}
\vspace{-3mm}
\label{AoI}
\end{figure}

\noindent The presented MOKE measurements have been conducted using a wavelength of 635\,nm and an AoI of 45$^\circ $ if not specified otherwise. More details on the MOKE setup can be found in the supplemental material. \\
For the magnetic characterization of the samples, MOKE curves have been measured for different in-plane orientations $\alpha $ in LMOKE geometry as depicted in Figs. \ref{MOKE}(a,e) for samples 1 and 2, respectively. As shown in Figs. \ref{MOKE}(b,c,f,g) the coercivity and remanent magnetization are rather isotropic with regard to $\alpha $ in both samples, indicating that they are magnetically isotropic. However, the MOKE response of the untwinned Co sample 1 in Fig. \ref{MOKE}(d) shows a pronounced threefold angular dependence in magnetic saturation at an external magnetic field of 230\,mT. This threefold angular dependence is a contribution of the LCMOKE to the MOKE signal. The effect can also be seen in the exemplarily displayed MOKE curves in Fig. 2(a). Due to the LCMOKE, the spans of the MOKE curves in mdeg increase or decrease depending on the sample orientation. In the twinned sample 2, the threefold angular dependence is reduced due to the twinning in the Co layer (c.f. Fig. \ref{MOKE}(h)). Since the two Co(111) phases effectively produce a threefold angular dependence with a phase shift of $\Delta \alpha$=60$^\circ $, the two angular dependencies cancel each other out. \\
For a more detailed, quantitative analysis of the different MOKE components, the eight-directional method has been performed on both samples. As can be seen in Fig. \ref{8dir}(a), $\Phi _{M_L,M_L^3}$ and $\Phi _{M_T^3}$ show strong angular dependencies in sample 1 due to the LCMOKE and the TCMOKE, respectively. 
The span between maximum and minimum of the $\Phi _{M_L,M_L^3}$ threefold angular dependence (\textit{i.e.}, twice the amplitude of the threefold angular dependence of $\Phi _{M_L,M_L^3}$) amounts to 14.8\% of the $\Phi _{M_L,M_L^3}$ offset that is primarily governed by LMOKE. For the Kerr ellipticity at 406\,nm wavelength this ratio is increased to 29.6\% (see supplemental material).
The QMOKE components $\Phi _{M_LM_T}$ and $\Phi _{M_T^2-M_L^2}$ only show negligible angular dependencies which are also negligible in the measurement of the Kerr ellipticity at 635\,nm and the Kerr rotation at 406\,nm (see supplemental material). Due to the different stacking order of the layers in sample 2, the $\Phi _{M_L,M_L^3}$ offset is drastically increased in contrast to the anisotropic contribution of $\Phi _{M_L,M_L^3}$ since the latter cancels out due to the strong twinning in the Co(111) layer. For both samples, a small onefold contribution to the MOKE components can be observed. This contribution can be explained as a superimposed vicinal MOKE (VISMOKE) signal \cite{Gaerner.2024, Hamrle.2003} which originates from a slight substrate miscut (see supplemental material).  \\
The experimental data can be fitted using numerical simulations based on Yeh's transfer matrix formalism \cite{PochiYeh.1980}, similarly done as in Ref. ~\onlinecite{Gaerner.2024}. The simulation model uses the layer thicknesses and $\epsilon _d$ of all layers (see supplemental material) as well as the wavelength and AoI of the light as fixed parameters. The model is then fitted to the experimental data with the MO parameters of the Co layer ($K$, $G_s$, $2G_{44}$ and $\Delta H$) set as free parameters. Note that since the contributions of $K$, $H_{123}$ and $3H_{125}$ cannot be evaluated independently (c.f. Eq. (\ref{ML})), the isotropic contribution of the CMOKE is minimized in the fits by setting $3H_{125}=0$ so that $\Delta H = H_{123}$. In order to account for the onefold VISMOKE contributions, an optostructural contribution $\varepsilon _s$ is added to the permittivity tensor (see supplemental material). The fits are displayed as solid lines in Figs. \ref{8dir}(a,b). The contributions  $\Phi _{M_L,M_L^3}$, $\Phi _{M_LM_T}$ and $\Phi _{M_T^2-M_L^2}$ are described rather well by the fits. Only the offset of $\Phi _{M_T^3}$ cannot be simulated, just as also observed in the case of Ni(111) thin films \cite{Gaerner.2024, Silber2026}. The fitted MO parameters of the simulation are displayed in Tab. \ref{MOparameters}. 
\begin{table*}[]
\begin{tabular}{ccc}
\hline
MO parameter & sample 1 & sample 2 \\ \hline
K & $-0.562\pm 0.002 - (0.186\pm 0.002)i$ & $-0.461\pm 0.001 - (0.130\pm 0.001)i$ \\
$G_s=G_{11}-G_{12}$ & $0.006\pm 0.002 - (0.014\pm 0.002)i$ & $0.005\pm 0.001 - (0.011\pm 0.001)i$ \\
2G$_{44}$ & $0.009\pm 0.001 - (0.019\pm 0.001)i$ & $0.007\pm 0.001 - (0.011\pm 0.001)i$ \\
$\Delta $H & $0.016\pm 0.001 + (0.0002\pm 0.0006)i$ & $0.0022\pm 0.0003 - (0.0004\pm 0.0003)i$ \\ \hline
\end{tabular}
\caption{MO parameters of samples 1 and 2 for a wavelength of 635\,nm determined by Yeh's 4x4 fits.}
\label{MOparameters}
\end{table*}
With the twinning of 86.9\% in sample 2, the real part of $\Delta H$ is reduced by 86.3\%, showing that $\Delta H$ correlates directly with the degree of twinning. In contrast, the real part of $K$ (found in the LinMOKE term) decreased only slightly in the twinned sample. \\
In theory, a threefold angular dependence of the MOKE signal measured in LMOKE geometry can also stem from a not fully saturated sample exhibiting a threefold magneto-crystalline anisotropy which results in $\bm{M}$ not always being parallel to $\bm{H}$ when $\bm{H}$ is applied along different in-plane directions. As shown in Figs. \ref{MOKE}(b,c) our sample does not possess such an anisotropy.
However, if it did, the threefold angular dependence could stem from variations in LMOKE, i.e. the linear contribution to the offset of $\Phi _{M_L,M_L^3}$ (c.f. Eq. (\ref{ML})). This offset is $\propto $ $B_{s/p}$ and therefore zero for perpendicular incidence while the anisotropic threefold contribution to $\Phi _{M_L,M_L^3}$, stemming from LCMOKE, is $\propto $ $A_{s/p}$ and therefore has a finite value at perpendicular incidence. In order to show that the anisotropic contribution to $\Phi _{M_L,M_L^3}$ is rather stemming from LCMOKE than from LinMOKE, we analyze its dependence on the AoI. The AoI was varied between 0$^\circ $ (perpendicular incidence) and 45$^\circ $ while $\Phi _{M_L,M_L^3}$ and $\Phi _{M_T^3}$ were measured for sample 1 depending on the sample orientation $\alpha $. The MO parameters for the simulations of the data with AoIs below 45$^\circ $ were extracted from the fit of the simulation to the experimental data at 45$^\circ$ AoI (c.f. Tab. \ref{MOparameters}) and kept constant for the simulation of the other AoIs. Results for $\Phi _{M_L,M_L^3}$  are shown in Fig. \ref{AoI}(a). It becomes evident that the model using the MO parameters yielded for 45$^\circ $ can predict the dependence of $\Phi _{M_L,M_L^3}$ on the AoI well. The experimental data was also fitted by the fit function
\begin{align}
    C + A_3 \sin{(3(\alpha -\alpha _3))} + A_1 \sin{(\alpha - \alpha _1)}
\end{align}
to extract the offset C and the amplitude of the threefold angular dependence $A_3$. As displayed in Fig. \ref{AoI}(b), the measured offsets for the different AoIs match the offsets predicted by the simulation for both $\Phi _{M_L,M_L^3}$ and $\Phi _{M_T^3}$. Only the offset of $\Phi _{M_T^3}$ starts to deviate from the expected value of zero as the AoI increases. As displayed in Fig. \ref{AoI}(c), the absolute values of the measured amplitudes of the threefold angular dependencies slightly increase with AoI, as expected from the simulation. The threefold angular dependence of $\Phi _{M_L,M_L^3}$ persists for normal AoI as predicted by the prefactor A$_{s/p}$ in Eq. (\ref{ML}) and our model of CMOKE describes its dependence on the AoI well. This anisotropic contribution cannot stem from LinMOKE since the contribution $\propto M_L$ would vanish at normal incidence, the same as the offset does. \\
 \\
\noindent As shown, CMOKE is not unique to Ni(111) thin films, but can also be observed in Co(111). Just as in Ni(111), the strength of the CMOKE anisotropy amplitude, or respectively the effective macroscopic size of the MO parameter $\Delta H$ (when averaging over all structural domains), varies with the degree of twinning in the Co layer. In contrast to LinMOKE contributions, the size of the CMOKE stays relatively constant with varying AoI, demonstrating that it can be detected in setups with different AoIs. \\
Furthermore, as demonstrated with these Co(111) samples, CMOKE can be strongly present even when a sample possesses almost no QMOKE. Here, the strength of the CMOKE reached up to about 30\% of the LinMOKE signal at 45$^\circ $ AoI. Therefore, CMOKE can be utilized, e.g. to sense the in-plane magnetization at normal AoI, which is otherwise a common application of QMOKE. However, since QMOKE is even in $\bm{M}$, it can only sense the axis of the in-plane magnetization while CMOKE can also sense its direction. Therefore, CMOKE provides a clear advantage for the detection of in-plane magnetization compared to QMOKE. \\

\appendix
\section*{Supplemental Material}
\noindent See the supplemental materials including Refs. \onlinecite{Kehlberger.2015, Muglich.2016, Keller.2002, Bjorck.2007, Parratt.1954, Johnson.1972, Johnson.1974, Powell.1970, Palik.1997} for additional information on QMOKE components and the used coordinate system, further details on the sample preparation and characterization as well as eight-directional measurements of the Kerr ellipticity and rotation at 406\,nm wavelength.
\section*{Acknowledgements}
\noindent J.H. acknowledge  financial support by the FerrMion project of the Czech Ministry of Education, co-funded by the EU, Project No. CZ.02.01.01/00/22\_008/0004591. We thank G. Reiss for making available laboratory equipment.

\section*{Data availability}
\noindent All data supporting the findings of this study are available upon reasonable request.

\nocite{*}
\bibliography{aipsamp}

\end{document}